\documentclass[twoside]{article}

\usepackage{Proc_NTSE}

\newcommand{\eg}{\textit{e.g.}}  

\newcommand{\ie}{\textit{i.e.}}

\newcommand{\be}{\begin {equation}}
\newcommand{\ee}{\end{equation}}
\newcommand{\bi}{\begin{itemize}}
\newcommand{\ei}{\end{itemize}}
\newcommand{\bea}{\begin {eqnarray}}
\newcommand{\eea}{\end{eqnarray}}
\newcommand{\braket}[2]{\bra{#1}\,#2\rangle} 
\newcommand{\bra}[1]{\langle\,#1\,|}          
\newcommand{\ket}[1]{|\,#1\,\rangle}          

\newcommand{\LCm}{{\scriptscriptstyle -}} 
\newcommand{\LCp}{{\scriptscriptstyle +}}

\begin{document}
\thispagestyle{plain}


\begin{center}
{\Large \bf \strut
 Non-Perturbative Time-Dependent Quantum Field Evolution
\strut}\\
\vspace{10mm}
{\large \bf 
Xingbo Zhao$^{a}$, Anton Ilderton$^{b}$, Pieter Maris$^{a}$ \\ and  James P. Vary$^{a}$}
\end{center}

\noindent{
\small $^a$\it Department of Physics and Astronomy, Iowa State University, Ames, Iowa 50011, USA} \\
{\small $^b$\it Department of Applied Physics, Chalmers University of Technology, SE-412 96 Gothenburg, Sweden} \\

\markboth{Xingbo Zhao, Anton Ilderton, Pieter Maris and James P. Vary}
{Non-Perturbative Time-Dependent Quantum Field Evolution}





%

\begin{abstract}
  We present a nonperturbative, first-principles numerical approach for time-dependent problems in the framework of quantum field theory. In this approach the time evolution of quantum field systems is treated in real time and at the amplitude level. As a test application, we apply this method to QED and study photon emission from an electron in a strong, time-dependent external field. Coherent superposition of electron acceleration and photon emission is observed in the nonperturbative regime.
  \\[\baselineskip]
\end{abstract}


%
\section{Introduction}

Solving time-dependent problems in quantum field theory is desired for a wide range of applications. One important area is in scattering processes.
Simulating scattering processes as time-dependent processes at the amplitude level opens up possibilities for handling complicated scenarios from first-principles. Typical examples include: 1) the asymptotic states cannot be well-defined. For example, long range forces exist between the colliding particles; another example is parton collisions in the deconfined medium created in relativistic heavy-ion collisions. 
2) The scattering processes occur in the presence of time-dependent background fields, which are typically encountered in strong field laser physics as well as in relativistic heavy-ion physics. In the former case, time-dependent electromagnetic fields are provided by laser beams, and in the latter case, colliding nuclei create strong and time-dependent (color-)electromagnetic fields. 3) One is interested in the explicit time evolution of quantum field amplitudes during scattering processes, which could shed light on, \eg, the mechanism of hadronization in QCD. 

To address time-dependent processes at the amplitude level, one first needs a stationary state description for stable particles
participating in the time-dependent process in terms of quantum field amplitudes. This was achieved by the previously constructed
Basis Light-front Quantization (BLFQ)~\cite{Vary:2009gt,Honkanen:2010rc}. The BLFQ adopts the light-front quantization and the
Hamiltonian framework, see~\cite{Brodsky:1997de} for a review on the light-front dynamics. It solves for the (boost-invariant)
light-front amplitudes for both bound states and scattering states by diagonalizing the light-front Hamiltonian of the quantum
field system. Recently, the efforts of applying BLFQ to positronium systems have been initiated~\cite{Wiecki:2013xx,Li:2013xx}.

In this paper, we introduce an extension of the BLFQ to the time-dependent regime, which is called the {\it time-dependent} Basis Light-Front Quantization (tBLFQ)~\cite{Zhao:2013cma,Zhao:2013jia}. Based on the stationary amplitudes obtained in BLFQ, tBLFQ evaluates the time-evolution of quantum field configurations by explicitly solving the time-dependent Schr\"odinger equation.
This approach provides a natural framework for addressing scattering problems from a time-dependent perspective. 

In this work, we illustrate tBLFQ through an application to strong field laser physics. Specifically, we study the ``nonlinear Compton scattering''(nCs) process~\cite{Nikishov:1963,DiPiazza:2011tq}, in which an electron is excited by a background laser field and emits a photon. This paper is organized as follows: in Sec.~\ref{sec:background}, we introduce our model for the background laser field; in Sec.~\ref{sec:tBLFQ} we discuss the formalism of tBLFQ; in Sec.~\ref{sec:BLFQ} we give a brief review on BLFQ which is employed to construct basis states for tBLFQ; in Sec.~\ref{sec:results} we conduct a sample calculation for the nCs process and present the numerical results. Finally we conclude and provide our outlook for future work in Sec.~\ref{sec:conclusion}.

\section{Background Field}
\label{sec:background}
We model the laser background as a classical field (\ie, we neglect back reaction on the laser). We consider a longitudinal periodic electric field pointing in the 3-direction with profile,
 \begin{align}
 E^3(x^+,x^-)=-E^3_0 \sin{(l_\LCm x^-)}\Theta(x^+)\Theta(\Delta x^+-x^+)\  ,
 \label{eq:laser_profile_electric_field}
 \end{align}
where $E^3_0$ is the peak amplitude and $l_\LCm$ is the frequency. The theta functions impose a finite light-front time duration on the field. An appropriate gauge potential is
\begin{align}
\mathcal{A}^-(x^+,x^-)=\frac{E^3_0}{l_\LCm} \cos{(l_\LCm x^-)}\Theta(x^+)\Theta(\Delta x^+-x^+)\ .
\label{eq:laser_profile}
\end{align}
The dependence on $x^+$ and $x^-$ makes this particularly suitable to a light-front treatment. 
\section{Quantum Evolution}
\label{sec:tBLFQ}

In tBLFQ, we calculate the evolution of quantum field configurations at the amplitude level.
For the nCs process, the Hamiltonian $P^-$ contains two parts, $P^-_\text{QED}$ which is the {\it full}, interacting, light-front Hamiltonian of QED, and interactions $V(x^\LCp)$ introduced by the external field, so
\begin{align}
	\label{H_interact}
	P^-(x^+)=P^-_\text{QED}+V(x^+) \;.
\end{align}

Both the QED Hamiltonian $P^-_{QED}$ and the external field interaction $V(x^\LCp)$ may induce transitions on the quantum field amplitudes over time. In the nCs process, we are, however, mostly interested in transitions induced by the external field $V(x^\LCp)$. Therefore, we adopt an interaction picture, in which the light-front QED Hamiltonian $P^-_{QED}$ serve as the ``main'' part of the Hamiltonian and the external field interaction $V(x^\LCp)$ as the ``interaction'' part. In this interaction picture, the quantum field amplitude evolves according to,
\be\label{Schro-int}
	i\frac{\partial}{\partial x^+}\ket{\psi;x^+}_I= \frac{1}{2}V_I(x^+)\ket{\psi;x^+}_I \;,
\ee
where $\ket{\psi;x^+}_I=e^{iP^-_{QED}x^+/2}\ket{\psi;x^+}$ is the quantum field amplitude in the interaction picture, and the ``interaction Hamiltonian in the interaction picture'' $V_I$ evolves in time according to
\be\label{V-int}
	V_I(x^+) = e^{\tfrac{i}{2}P^-_\text{QED}x^+}V(x^+)e^{-\tfrac{i}{2}P^-_\text{QED}x^+} \;.
\ee
The solution to~(\ref{Schro-int}) can be formally written in terms of a time-ordered ($ \mathcal{T}_+$) series, as:
\begin{align}
	\label{i_evolve}
	\ket{\psi;x^+}_I  &= \mathcal{T}_\LCp e^{-\tfrac{i}{2}\int\limits_{x^+_0}^{x^+} V_I\, dx^+}\ket{\psi;x^+_0}_I \;,
\end{align}
where $\ket{\psi;x^+_0}_I$ is the initial quantum field amplitude at light-front time $x^+_0$. In the nCs process, this initial state corresponds to a single physical electron.

To implement this solution numerically, we need to specify a basis for the quantum field amplitudes $\ket{\psi;x^+}_I$ as well as the external field interaction $V_I(x^+)$. Eq.~(\ref{V-int}) suggests that the most convenient basis is what comprises the eigenstates of the light-front QED Hamiltonian, $P^-_{QED}$. We denote this basis as $\ket{\beta}$, which can be found by solving the eigenvalue problem for $P^-_{QED}$, 
\begin{align}
    \label{eq:hami_QED_eigeneq}
    P^-_{QED}\ket{\beta}=P^-_\beta\ket{\beta}\ ,
\end{align}
where $P^-_\beta$ is the eigenvalue (light-front energy) for the eigenstate $\ket{\beta}$. In tBLFQ, we employ the previously constructed BLFQ~\cite{Vary:2009gt,Honkanen:2010rc} to solve this eigenvalue problem. More details will be shown in the next section. 

In terms of the basis states $\ket{\beta}$, the quantum field state $\ket{\psi;x^+}$ is represented as,
\be\label{ket_int-expand}
	\ket{\psi;x^+}_I  = \sum_\beta c_\beta(x^+) \ket{\beta}\ ,
\ee
where $c_\beta(x^+)=\braket{\beta}{\psi;x^+}$ is the amplitude in the basis $\ket{\beta}$. The initial state in the nCs process -- a physical electron -- is an eigenstate of $P^-_{QED}$ and thus can be trivially expressed in this basis.


With both the quantum field configuration and the interaction term in the Hamiltonian represented in the basis $\ket{\beta}$, Eq.~(\ref{i_evolve}) can be realized as a series of matrix-vector multiplications acting on an initial state vector. To make the numerical calculation feasible, in this step, we make two approximations: ``time-step discretization'' and ``basis truncation'': the former is to decompose the time-evolution operator in (\ref{i_evolve}) into small but finite steps in light-front time $x^+$, with the step size $\delta x^+$, 
%
\be
\label{sol_wave-eq_i_discrete}
\mathcal{T}_\LCp e^{-\frac{i}{2}\int\limits_{x^+_0}^{x^+} V_I\, dx^+}\ket{\psi(x^+_0)}_I\rightarrow \big[1-\tfrac{i}{2}V_I(x^+_{n})\delta x^+\big] \cdots \big[1-\tfrac{i}{2}V_I(x^+_{1})\delta x^+\big]\ket{\psi(x^+_0)}_I \;,
\ee
and the latter is to keep the basis dimensionality finite.  In tBLFQ, the basis truncation is performed in the basis state ($\ket{\beta}$) construction stage in BLFQ, which will be introduced in the next section. ``Basis truncation" and ``time-step discretization" are the only two approximations in tBLFQ.

\section{Basis Construction}
\label{sec:BLFQ}

In this section we present a brief review of BLFQ~\cite{Vary:2009gt,Honkanen:2010rc} and explain the procedure of constructing the tBLFQ basis $\ket{\beta}$ through solving the eigenvalue problem of $P^-_{QED}$ in BLFQ. For more details, see Ref.~\cite{Zhao:2013cma}.



Since quantum field systems generally have large numbers of degrees-of-freedom, to mitigate the computational burden, it is important to choose an efficient basis for the eigenvalue problem. The idea in BLFQ is that an efficient basis should capture the symmetries of the underlying dynamics, so that in such a basis the Hamiltonian exhibits a block-diagonal structure. Each block is associated with a group of quantum numbers corresponding to the symmetries captured by the basis. Thus, the Hamiltonian can be diagonalized block by block, and in practice, we can selectively diagonalize only those blocks with desired quantum numbers.


Specifically for the light-front QED Hamiltonian $P^-_\text{QED}$, the BLFQ basis, denoted as $\ket{\alpha}$, captures the following three symmetries: 1) Translational symmetry in the $x^-$ direction; 2) Rotational symmetry in the transverse plane; 3) Lepton number conservation. These three symmetries correspond to three conserved operators. These are the longitudinal momentum $P^\LCp$, longitudinal projection of angular momentum $J^3$, and charge, or net fermion number $Q$, respectively. The BLFQ basis states $\ket{\alpha}$ are chosen to be the eigenstates of these three operators: $\{P^+,J^3,Q\}\ket{\alpha}=\{2\pi K/L,M_j,N_f\}\ket{\alpha}$ ($L$ is the length of the longitudinal ``box'' in which we embed our system, see below). These eigenvalues divide the BLFQ basis states $\ket{\alpha}$ into multiple ``segments''. Each segment consists of the basis states $\ket{\alpha}$ sharing the same group of eigenvalues.


BLFQ basis states $\ket{\alpha}$ are constructed in terms of a Fock sector expansion. 
Each Fock particle has helicity, longitudinal momentum $p^+=2\pi k/L$ (the sum of $k$, in each state, must equal $K$, which implicitly imposes basis truncation on the longitudinal degree of freedom), and two transverse degrees of freedom. The latter are described in terms of the radial quantum number, $n$, and the angular quantum number, $m$, of the eigenstates of a 2D-harmonic oscillator (2D-HO). This choice of basis, motivated by applications to QCD, is suitable for describing the confining interaction. This choice is supported by the success of the AdS/QCD approach to hadron spectroscopy~\cite{deTeramond:2008ht}, where a similar basis is adopted. 

In summary, a complete specification of a BLFQ basis requires 1) the segment specifiers $K,M_j$, and $N_f$, 2) two truncation parameters, namely the choice of which Fock sectors to retain, and the transverse truncation parameter $N_\text{max}$, the maximum total number of oscillator quanta $2n+|m|$ for the Fock states and 3) the ``box length'' $L$ in the longitudinal direction and a scale parameter $b=\sqrt{M\Omega}$ for the 2D-HO wave functions, where $M$ and $\Omega$ are the mass and frequency of the 2D-HO. 

Specifically for the nCs processes, the initial state is a single physical electron. Since the external field interaction $V(x^+)$ conserves net fermion number and does not excite transverse degrees of freedom, we only need to prepare eigenstates of $P^-_{QED}$ in segments of different $K$.
The $K$'s in these segments are equally spaced by the longitudinal momentum quantum number of the background field $k_{las}=Ll_-/\pi$. In each segment, we truncate the Fock sectors to the lowest two sectors, $\ket{e}$ and $\ket{e\gamma}$. We take $L=2\pi$\,MeV$^{-1}$, so the value of $k(K)$ can be read as the longitudinal momentum in units of MeV, and $b=0.511$\,MeV, which matches the natural scale in QED set by the physical electron mass $m_e$.

In our constructed BLFQ basis $\ket{\alpha}$, $P^-_{QED}$ manifests as a block-diagonal numerical matrix. Each block is associated with a distinct $K$. Then, upon diagonalizing $P^-_{QED}$ block by block (see~\cite{Karmanov:2008br} for sector-dependent renormalization, and~\cite{Chabysheva:2009ez,Zhao:2013xx} for applications), we obtain the eigenvalues $P^-_\beta$ and the associated eigenstates $\ket{\beta}$ in the basis
$\ket{\alpha}$,
\begin{align}\label{phys-expand}
    \ket{\beta}=\sum_\alpha \ket{\alpha} \braket{\alpha}{\beta} \;.
\end{align}
These eigenstates $\ket{\beta}$ are the basis states used in tBLFQ. For each basis state, the invariant mass $M_\beta$ relates to its eigenvalue $P^-_\beta$ as: $M^2_\beta=P^+_\beta P^-_\beta-P^{2}_{\perp,\beta}$, where $P^+_\beta$ and $P_{\perp,\beta}$ are the longitudinal and transverse momentum for $\ket{\beta}$, respectively. The ground states (with the lowest invariant mass in each segment) are interpreted as the physical electron states and the excited states are interpreted as the electron-photon scattering states.

\section{Numerical Results}
\label{sec:results}

\begin{figure}[!tbh]
\centering
\includegraphics[width=0.8\textwidth]{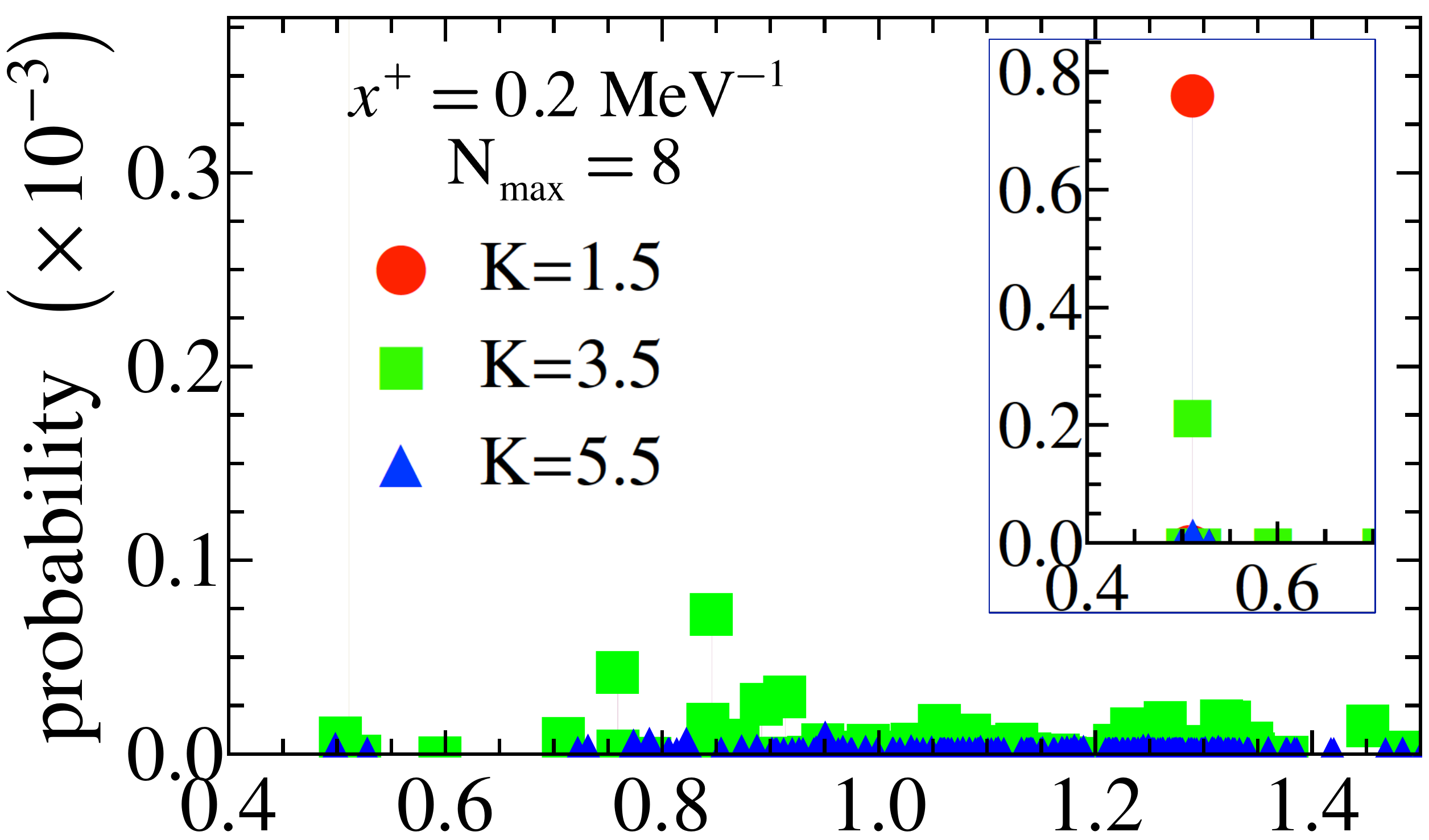}
\includegraphics[width=0.8\textwidth]{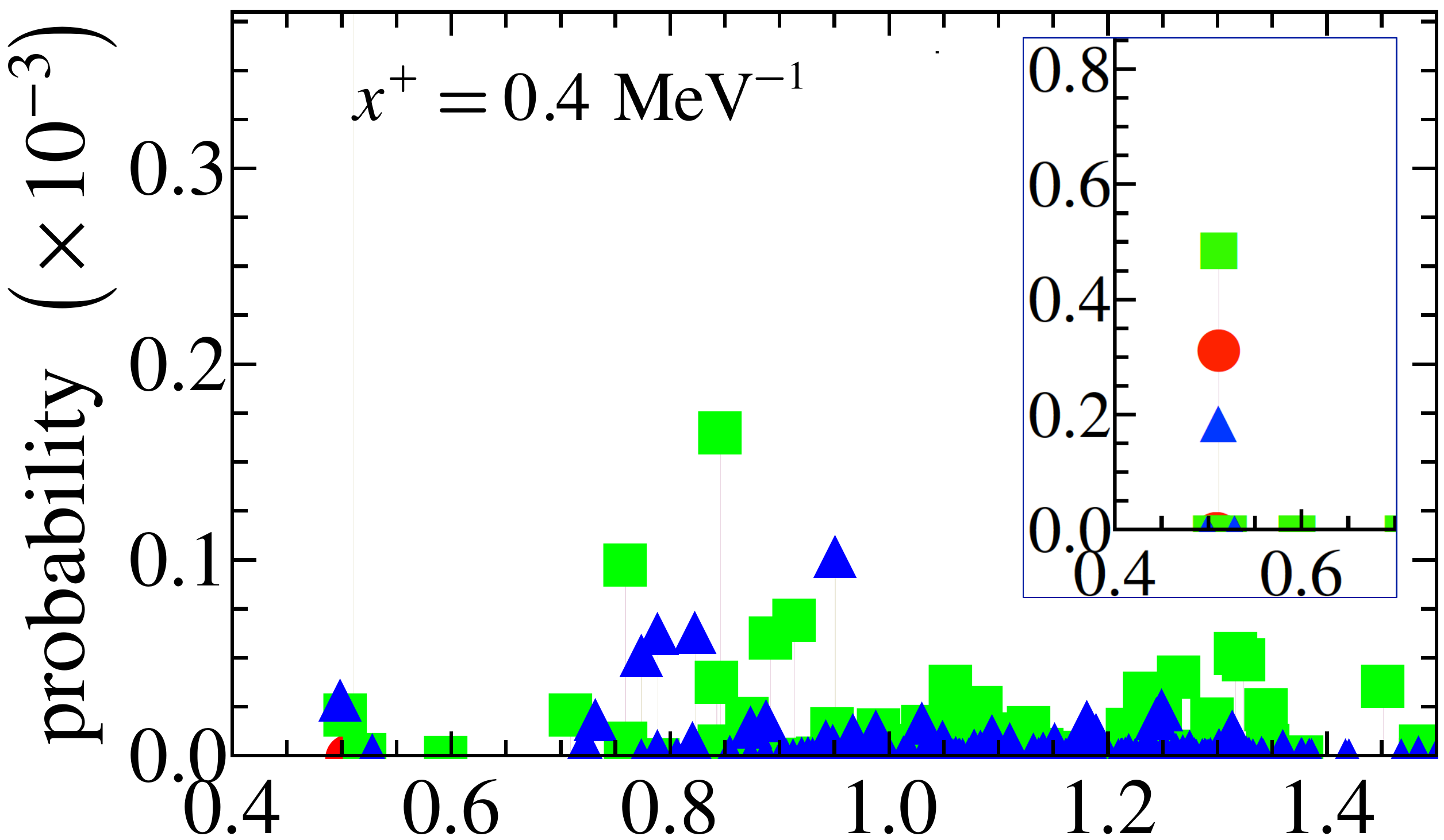}
\includegraphics[width=0.8\textwidth]{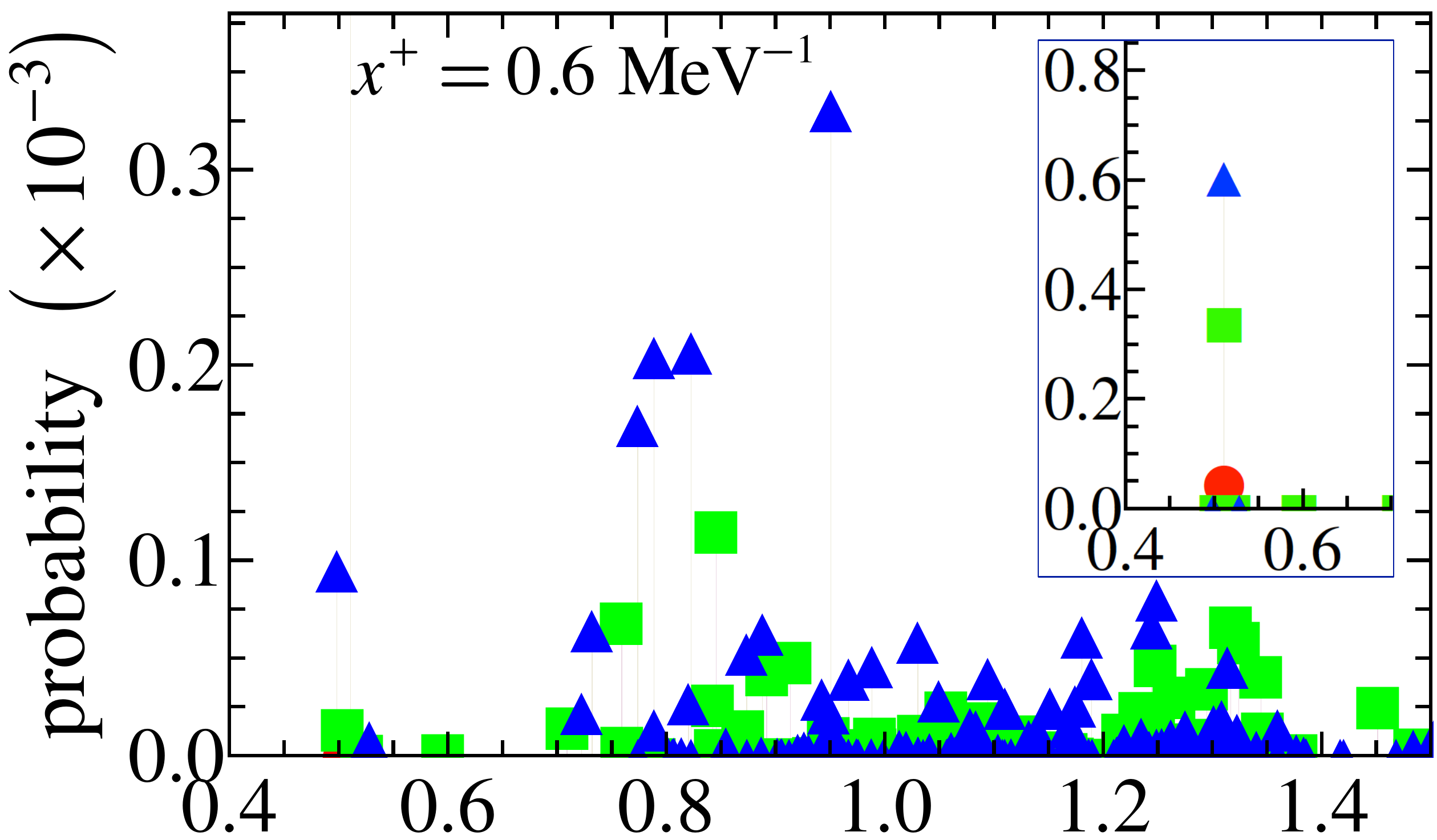}
\caption{\label{fig:state_evol_np} (Color online) Time evolution of the system at (from top to bottom) $x^+= 0.2,0.4,0.6$\,MeV$^{-1}$, with the background field switched on at $x^+$=0. Each dot represents a tBLFQ basis state $\ket{\beta}$, an eigenstate of QED. Horizontal axis: the invariant mass of the state $M_{\beta}$. Vertical axis: the probability of finding the state $\ket{\beta}$ in units of $10^{-3}$. The inset panels show, at normal scale, the (much larger) probabilities of finding the single physical electron states (in the $K{=}1.5, 3.5, 5.5$ segments), with invariant mass $M_\beta=m_e$.} 
\end{figure}

In this section we carry out a sample numerical calculation for the nCs process.
A basis consisting of three segments with $K{=}\{K_i, K_i{+}k_\text{las}, K_i{+}2k_\text{las}\}$ is chosen for this calculation. In each segment we retain {\it both} the single electron (ground) and electron-photon (excited) state(s). The initial state for our process is a single (ground state) electron in the $K{=}K_i$ segment. This basis allows for the ground state to be excited twice by the background (from the segment with $K$=$K_i$ through to segment with $K_i$+2$k_\text{las}$). In this calculation, we take $K_i{=}1.5$ and $N_\text{max}{=}8$, with $a_0$=10, $k_\text{las}{=}2$ and $L{=}2\pi$\,MeV$^{-1}$. We present the evolution of the electron system in Fig.~\ref{fig:state_evol_np}, at increasing (top to bottom) light-front times.
As time evolves, Fig.~\ref{fig:state_evol_np} shows how the background causes transitions from the ground state in the $K{=}1.5$ segment to other eigenstates of $P^\LCm_\text{QED}$. Both the single electron states and electron-photon states are populated; the former represent the acceleration of the electron by the background, while the latter represent the process of radiation. At times $x^+{=}0.2$\,MeV$^{-1}$, the single electron state in the $K{=}3.5$ segment becomes populated while the probability for finding the initial state begins to drop. The populated electron-photon states begin forming a peak structure. The location of the peak is around the invariant mass of 0.8\,MeV, roughly consistent with the expected value of $M_\text{pk1}{=}\sqrt{P^-_i(K_i+k_\text{las})}{=}0.78$\,MeV, where $P^-_i=\frac{m^2_e}{K_i}\sim 0.17$\,MeV is the light-front energy of the initial single electron state with $K{=}3.5$.

As time evolves, the probability of finding the electron in its initial (ground) state continues to decrease. Single electron states with successively higher $K$ dominate the system. It also becomes possible to find electron-photon states of higher $K$ and invariant mass, following the absorption of more energy from the background field as time evolves (see the second and third rows of Fig.~\ref{fig:state_evol_np}).

\begin{figure}[t]
\centering
\includegraphics[width=0.8\textwidth]{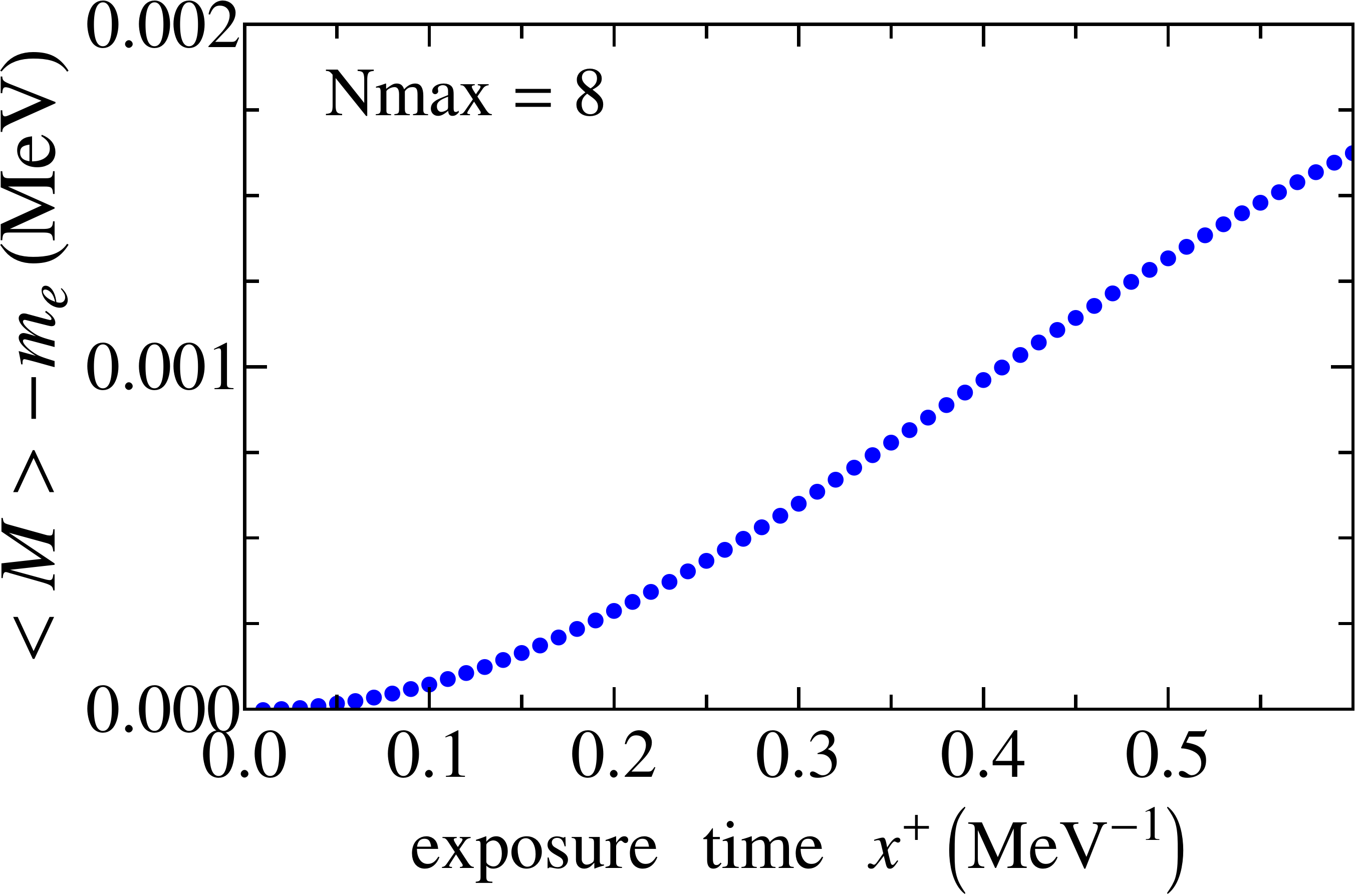}
\caption{\label{fig:invmass_evol_np} (Color online) The average invariant mass of the electron-photon system evolves as a function of exposure time. The vertical axis is the difference between the average invariant mass of the electron-photon system and that of a single physical electron.}
\end{figure}

As the state $\ket{\psi;x^\LCp}$ encodes all the information of the system, other observables can be constructed out of $\ket{\psi;x^\LCp}$. As an example, the evolution of the average invariant mass $\langle M\rangle\equiv\sum_\beta M_\beta\bra{\beta}\psi\rangle^2$ of the system as a function of exposure time is displayed in Fig.~\ref{fig:invmass_evol_np}. The approximately linear increase of the invariant mass up to an exposure time of 0.6\,MeV$^{-1}$ indicates the fact that photons are created as the background field pumps energy into the system.

\section{Conclusion and Outlook}
\label{sec:conclusion}
In this paper, we review a recently constructed nonperturbative framework for time-dependent problems in quantum field theory. It is called ``time-dependent BLFQ" (tBLFQ). Adopting the light-front dynamics and Hamiltonian formalism, tBLFQ provides the evolution of quantum field amplitudes through the light-front Schr\"odinger equation. Given the light-front Hamiltonian of the system, and an initial state as input, the quantum field amplitudes of the system at any subsequent time can be evaluated. The entire calculation is performed nonperturbatively with basis truncation and time-step discretization being the only two approximations. 

As a generic method for time-dependent problems in quantum field theory, the tBLFQ method can be applied to both first-principles and effective Hamiltonians, where the time-dependence arises either from the background fields or from using non-stationary initial states. In this work we apply the tBLFQ method to strong field laser physics and specifically study the nonlinear Compton scattering process, in which an electron is accelerated by a background field and emits a photon. The numerical calculation reveals a coherent superposition of electron acceleration and photon-emission processes in the nonperturbative regime.


As a next step, we plan to apply this method to relativistic heavy-ion collisions in which the medium of the colliding nuclei can be modeled as a dissipative background field. For example, the energy loss of the produced quark and gluon jets in this evolving background field can be predicted. Another application is the hadronization process in QCD.

We acknowledge valuable discussions with K. Tuchin, H. Honkanen, S. J. Brodsky, P. Hoyer, P. Wiecki and Y. Li. This work was supported in part by the Department of Energy under Grant Nos. DE-FG02-87ER40371 and DESC0008485 (SciDAC-3/NUCLEI) and by the National Science Foundation under Grant No. PHY-0904782. A.~I.\ is supported by the Swedish Research Council, contract 2011-4221.


\end{document}